# Visualization of dark excitons in semiconductor monolayers for high-sensitivity strain sensing

*Saroj B. Chand[1], John M. Woods[1], Enrique Mejia[1], Takashi Taniguchi[2], Kenji Watanabe[3], and Gabriele Grosso[1,4]\**

1. Photonics Initiative, Advanced Science Research Center, City University of New York, New York, NY 10031, USA
2. International Center for Materials Nanoarchitectonics, National Institute for Materials Science, 1-1 Namiki, Tsukuba 305-0044, Japan
3. Research Center for Functional Materials, National Institute for Materials Science, 1-1 Namiki, Tsukuba 305-0044, Japan
4. Physics Program, Graduate Center, City University of New York, New York, NY 10016, USA

*Corresponding Author e-mail: ggrosso@gc.cuny.edu

ABSTRACT

Transition metal dichalcogenides (TMDs) are layered materials that have a semiconducting phase with many advantageous optoelectronic properties, including tightly bound excitons and spin-valley locking. In Tungsten-based TMDs, spin and momentum forbidden transitions give rise to dark excitons that typically are optically inaccessible but represent the lowest excitonic states of the system. Dark excitons can deeply affect transport, dynamics and coherence of bright excitons, hampering device performance. Therefore, it is crucial to create conditions in which these excitonic states can be visualized and controlled. Here, we show that compressive strain in $WS_2$ enables phonon scattering of photoexcited electrons between momentum valleys, enhancing the formation of dark intervalley excitons. We show that the emission and spectral properties of momentum-forbidden excitons are accessible and strongly depend on the local strain environment that modifies the band alignment. This mechanism is further exploited for strain sensing in two-dimensional semiconductors revealing a gauge factor exceeding $10^4$.

Semiconducting monolayers of TMDs combine promising optoelectronic properties with physical qualities characteristic of two-dimensional (2D) materials, such as high flexibility, few-atom thickness, easy transfer and assembly. Their optical response is dominated by excitons,[1] Coulomb-bound electron-hole pairs, with large binding energy and peculiar spin-valley coherence properties.[2] This unique combination of features offers the possibility to tailor their optical and electronic attributes by different means, including stacking of different layers[3] and strain engineering.[4] Due to their strong in-plane covalent bonding, TMDs can sustain strain of a few percent with dramatic changes in the electronic band diagram[5] and electron-phonon coupling.[6] Strain can strongly affect the exciton emission energy and linewidth,[7] and it is an efficient method to modulate the exciton potential landscape and achieve exciton funneling[8,9] as well as single-exciton localization.[10,11]

The electronic band structure of monolayer $WS_2$, shown in Fig.1a, gives rise to several excitonic resonances.[12,13] In the monolayer limit, $WS_2$ shows valence band (VB) maxima at the *Γ* and *K* high-symmetry points, and conduction band (CB) minima at *K* and *Λ*. The broken inversion symmetry of the system generates non-equivalent points at *K'* and *Λ'*.[14] This structure allows for the formation of two bright intravalley excitons, *KK* and *K'K'* with opposite spin, and their complexes.[15] Four indirect intervalley excitons *KΛ, KK', ΓK* and *ΓΛ* can also be formed, but due to their large momentum that cannot be directly carried by photons, they are called momentum-forbidden dark excitons. Note that spin-forbidden excitons can also form because of the large spin-orbit coupling that splits the minima of the conduction bands at *K* and *K'* resulting in one spin-allowed and one spin-forbidden transition.[16] Although the *KK* and *K'K'* are the only optically-accessible excitons in this system, they are not the excitonic ground state which is instead represented by the momentum-forbidden *KΛ* exciton which possesses larger binding energy and slightly longer lifetime, in the order of a few picoseconds.[17] Being the ground state of the system, *KΛ* excitons can deeply affect overall exciton dynamics, coherence and linewidth,[18] and controlling them is crucial for optoelectronic applications. Recently, a theoretical work predicted that, in the presence of compressive strain, momentum-forbidden *KΛ* dark excitons appear in the optical spectrum.[19] Although the 2D nature of TMDs allows them to sustain high tensile strain,[20] they have a proclivity to buckle under compressive strain.[21] Therefore, creating compressive strain in practical experiments, especially at low temperatures, represents a challenge. Here, we investigate the optical response of a monolayer $WS_2$ in which strain is impressed during the transfer process onto a hexagonal boron nitride (hBN) substrate. We show that dark excitons appear around compressive strain pockets with emission intensity and energy that are correlated to the local strain magnitude.

We first discuss how strain differently impacts the electronic bands at the high symmetry points in the Brillouin zone. In the unstrained case, our simulations based on density functional theory (DFT) indicate that the minimum of the conduction band at the *K* valley lies below the one at *Λ* with $\Delta E^{KΛ}=E^Λ-E^K =28$ meV, in agreement with previous studies.[19,22,23] When compressive strain is applied, both conduction band minima shift, albeit at different rates due to their orbital compositions.[24] When we account for movement of the VB maximum and CB minima at $K$ and $\Lambda$ under strain, we find that $\Delta E^{K\Lambda}$ changes at a rate of 213 meV/% (see Fig.S1), which is consistent with previously reported values.[19] Since the effect of strain on the spin-orbit coupling[25] and the curvature of the bands is small,[19] we can neglect the variation of orbital gap and exciton binding energy. See Table S1 for a summary of the strain-induced effects on the effective masses and energy at high symmetry points. Note that in this work we will focus on biaxial strain which has roughly doubled effects when compared to the uniaxial case[7] and more accurately reflects the strain loading imparted by the transfer process.

Migration of carriers across different valleys can occur via scattering with phonons by conserving energy and momentum. Fig.1b shows the phonon dispersion in an unstrained monolayer of $WS_2$. Monolayer TMDs have a $D_{3h}$ symmetry and nine phonon branches,[19] but only the LA, TA, E'(LO and TO), and $A_1$ modes are relevant for scattering with electrons and holes.[22,26] In $WS_2$, the dominant modes are LA phonons in the conduction band and TA phonons in the valence band.[22,26] Therefore, the electron scattering from $K$ to $\Lambda$ is mainly assisted by LA phonons with momentum $\Lambda$ that have energy of 18.8 meV (Table S2). Our simulations indicate that $\Delta E^{K\Lambda}$ has a similar value at approximately -0.04% compressive strain. The involvement of LA phonons in the formation of momentum-forbidden $K\Lambda$ excitons has been previously demonstrated with near-resonance Raman spectroscopy.[27] The electron scattering from $K$ to $K'$ is also assisted by acoustic phonons but with weaker electron-phonon coupling.[22,28] Note that, differently from the electronic dispersion, phononic modes do not change considerably in this range of strain,[7] and their energy and momentum can be considered constant.

Fig.1c shows the phonon-assisted process that enables the formation and radiative emission of dark excitons in $WS_2$. In the unstrained case, upon optical excitation, electrons are excited from the CB to the VB at the $K$ valley. This initial carrier population thermalizes and gives rise to the direct $KK$ excitons that radiatively recombine. When compressive strain is applied, $\Delta E^{K\Lambda}$ gradually decreases until one LA phonon can cover the energy and momentum mismatch between $K$ and the $\Lambda$ valleys, enabling the scattering of electrons and formation of $K\Lambda$ excitons. $K\Lambda$ excitons can then recombine radiatively by emitting phonons and scattering to virtual states within the light cone.[29] Fig.1d and Fig.1e show the emission spectra of a monolayer $WS_2$ at 77 K for unstrained and compressively strained cases, respectively. In the pristine sample (Fig.1d), the $KK$ exciton peak at $E^{KK}_{exc}$=2.076 eV is followed by two low energy peaks attributed to singlet and triplet trionic exciton complexes.[15] In the presence of compressive strain, the $K\Lambda$ exciton peak appears in the spectrum at $E^{K\Lambda}_{exc}$=1.913 eV. The energy separation between the $KK$ and $K\Lambda$ excitons ($\Delta E^{K\Lambda}_{exc}$) depends on the strain environment and temperature,[30] and we find $\Delta E^{K\Lambda}_{exc}$=163 meV at T=77 K.

To investigate the correlation between the $K\Lambda$ exciton and strain, we analyze the strain landscape of our samples. We deposit the mechanically exfoliated $WS_2$ monolayers onto hBN substrates with a dry transfer technique by using a polydimethylsiloxane (PDMS) stamp (see Methods). The interplay between the van der Waals forces of the hBN substrate and PDMS generates numerous wrinkles and nanobubbles (see Fig.S2 & Fig.S3). This crumpling effect is only observed when transferring onto hBN and not SiO$_2$ which has a smaller adhesion energy.[31] A typical topography of the sample, obtained by atomic force microscopy (AFM), is illustrated in Fig.2a and shows a surface with sunken and elevated regions. The peculiar structure of the monolayer results in a rich strain environment with a number of tensile and compressive strain regions as shown in Fig.2b, in which strain is modeled with the method described in ref.[32] (more details in SI). Based on this calculation, we assume that strain in our samples can range from -0.1% to +0.3%. Despite the uneven surface, the PL emission (Fig.2c) is relatively homogeneous. Fig.2d shows the PL emission from the same area measured using a long-pass filter in order to only collect photons with energy below 1.94 eV and cut out the emission from the *KK* exciton complexes. Several localized spots of *KΛ* excitons emission are visible across the sample.

To measure strain, we monitor across the sample the *KK* exciton energy ($E^{KK}_{exc}$) that is directly related to the variation of the bandgap. We perform hyperspectral imaging of the sample in an area of 6 μm x 6 μm with 100 nm-size steps and we extract $E^{KK}_{exc}$ and the intensity of the *KΛ* exciton ($I^{KΛ}_{exc}$) by Lorentzian fits (more details in Fig.S5). Fig.2e shows the variation of $E^{KK}_{exc}$ overlaid to the contour plot of $I^{KΛ}_{exc}$ in the same region. $E^{KK}_{exc}$ varies by 6 meV in the sampled area due to the presence of compressive and tensile strain. The *KΛ* exciton emission is localized around regions of compressive strain. Fig.2f shows the intensity of the *KΛ* exciton around a localized emission spot as a function of $E^{KK}_{exc}$. For low values of $E^{KK}_{exc}$ (corresponding to smaller applied compressive strain), the *KΛ* exciton is barely visible, and it quickly gets more intense when $E^{KK}_{exc}$ increases due to higher compressive strain; however, as $E^{KK}_{exc}$ continues to increase from further compressive strain, $I^{KΛ}_{exc}$ slowly decreases back towards zero. This brief window of dark exciton emission is explained by the narrow distribution of the phononic density of states of WS$_2$ at this energy,[33] the large dispersion of the LA phonons around the *Λ* point ($|\Delta E/\Delta K|$=24 meV/Å$^{-1}$, see inset in Fig.1b), and the relatively flat electronic band at the *Λ* point. The involvement of the *K→Λ* phononic scattering process is also confirmed by resonant excitation experiments. Fig.3a shows the PL map of a WS$_2$ flake at T=5 K excited at 532 nm. The laser signal is filtered out from the PL emission with a long pass filter at 550 nm and emission from both bright and dark excitons is collected. Fig.3b shows the PL emission from the same sample excited by a supercontinuum pulsed laser with a 2nm-band tunable spectral filter centered at 532 nm. In this case, a long pass filter at 640 nm (1.94 eV) is used in detection to collect only emission from the *KΛ* excitons that, as for the sample in Fig.2d, is localized around compressive strain regions. We then perform the same experiments by exciting the sample in resonance with the *KK* exciton. The emission pattern of the *KΛ* exciton obtained with non-resonant excitation (Fig.3b) is the same as the one obtained with resonant excitation on the *KK* exciton, as shown in Fig.3c. Moreover, the intensity of the *KΛ* emission is more intense in the case of resonant excitation. Interestingly, when the sample is excited below the *KK* exciton resonance at 632 nm (1.96 eV), Fig.3d, the *KΛ* emission pattern changes dramatically and the intensity vanishes. In this configuration, the sample is excited below the *KK* exciton resonance but above the *KΛ* exciton energy. The fact that the dark exciton emission almost vanishes suggests that the *K→Λ* scattering channel is the main mechanism for the formation of *KΛ* excitons.

To further investigate the role of phonons in the formation of *KΛ* excitons, we perform power dependent measurements at different temperatures. Fig.4a shows the variation of $I^{K\Lambda}_{exc}$ at different excitation powers at T=5, 60, 120, 180 K. At lower temperatures, $I^{K\Lambda}_{exc}$ shows a saturation behavior that is explained by the low density of states of LA phonons at this particular energy and momentum.[33] When the temperature increases, more phonons are available to promote the intervalley scattering process and $I^{K\Lambda}_{exc}$ increases almost linearly with the excitation power *P*. The increase of the phononic states is quantified by fitting the intensity curves with a power function $I(P)=AP^w$. The exponent *w* increases with temperature, from *w*=0.18 at T=5 K to *w*=0.77 at T=180 K. Although the efficiency of the *K→Λ* scattering channel increases with temperature, the latter also increases the interatomic distance generating an effect similar to the one of tensile strain. When temperature increases, the action of compressive strain on the band alignment is reduced and $\Delta E^{K\Lambda}$ increases. The spectral weight of *KK* and *KΛ* excitons as a function of temperature is shown in Fig.4b, indicating that the efficiency of the formation and emission process for *KΛ* excitons is temperature dependent and, in this strain condition, has a maximum around T=60 K. Fig.4c shows how the emission energy for both *KK* and *KΛ* excitons changes with temperature. While the *KK* exciton redshifts of approximately 40 meV when the temperature is raised from 5 K to 210 K, the *KΛ* exciton energy changes only by 15 meV. This observation supports the theoretical prediction of a faster shift of the electronic bands at the *K* point compared to the *Λ* point when the interatomic distance increases. Note that intervalley scattering influences both coherence and fluorescence lifetime of *KK* excitons.[6,18,34] Since phonon-assisted intervalley scattering plays an important role in the formation and recombination of *KΛ* excitons, their decay rate is expected to depend on strain and temperature that dramatically affect band alignment and thus intervalley scattering.

Due to the drastic change in the emission spectrum, the appearance of the *KΛ* exciton can be used as a high-sensitive mechanism to measure compressive strain.[19] To explore this possibility, we estimate the strain generated in a small region of interest and perform hyperspectral scans to extract the distribution of $E^{KK}_{exc}$ and $I^{K\Lambda}_{exc}$. This experiment is performed at T=60 K to maximize the formation of dark excitons. In the region of interest (see Fig.S7), strain has an asymmetric distribution ranging from -0.02% to +0.04%. Since $E^{KK}_{exc}$ scales linearly with strain, we can then estimate a zero-strain value for the emission energy of the *KK* exciton, $E^{KK,0}_{exc}$, assuming that the distribution of $E^{KK}_{exc}$ follows the one of strain (see Fig.S6). We obtain $E^{KK,0}_{exc}$=2.0852 eV at T=60 K. Due to technical challenges associated with measuring the strain-induced energy shift of excitons in TMDs, especially at cryogenic temperatures, there is no general agreement on how much the *KK* exciton energy shifts as a function of strain.[35] Experimental values for uniaxial strain range from 10 meV/% to 61 meV/% at room temperature, with the most common value of 45 meV/%. Considering that the strain in our samples stems from wrinkles and nanobubbles, it is safe to assume biaxial loading, therefore we use 90 meV/%. We can then estimate the strain magnitude from the experimental value of $E^{KK}_{exc}$: $\epsilon(\%)=(E^{KK}_{exc}-E^{KK,0}_{exc})/90$. From the emission spectra at different values of strain at T=60 K shown in Fig.5a, it is possible to discern a clear blueshift of $E^{KK}_{exc}$ and an increase (decrease) of the *KΛ* (*KK*) exciton intensity. Note that $E^{KΛ}_{exc}$ decreases as expected from the shift of the CB at the *Λ* valley. The intensity of both excitons, $I^{KK}_{exc}$ and $I^{KΛ}_{exc}$, show an exponential change (Fig. 5b). In these conditions, the *KΛ* exciton band lies below the *KK* exciton band (see Fig.1c) and their occupations are given by the Bose-Einstein distribution.[19] The linewidth of the *KK* excitons also increases as a function of compressive strain, confirming the gradual involvement of phonons (see Fig.S5c).[18] Finally we can estimate the optical gauge factor defined as $g_F = \frac{(I^S-I^0)/I^0}{\epsilon(\%)}$, where $I^S$ and $I^0$ are the intensities of the *KΛ* exciton with and without strain, respectively. Since $I^0$ would be null at zero strain, we use the smallest value of $I^{KΛ}_{exc}$ detected in this dataset. Fig.5c shows the optical gauge factor as a function of strain, indicating ultra-high sensitivity with $g_F$ values exceeding $10^4$.

In summary, we have reported on the experimental observation of momentum-forbidden *KΛ* excitons which are the excitonic ground state of monolayer $WS_2$. We have shown evidence that the formation and emission of *KΛ* excitons is related to compressive strain that activates a phonon-assisted intervalley scattering process. An ultra-sensitive optical strain sensing mechanism based on the brightening of dark excitons is also proposed indicating new routes to measure small variations of strain in two-dimensional semiconductors.

METHODS

**Sample preparation.** 2D materials are mechanically exfoliated using a standard scotch tape method and then transferred onto a Si/SiO2 substrate. We use hBN flakes of thickness around ~20nm as our bottom layer. Monolayers of WS$_2$ are initially exfoliated from the bulk crystals (from HQ Graphene) on a Polydimethylsiloxane (PMDS) stamp (X4 WF Film from Gel-Pak) and then moved onto the hBN substrate using a standard dry transfer technique. The interplay between WS$_2$ adhesion with the hBN substrate and the PDMS results in the crumbling of the monolayer once it is deposited on the hBN creating a rich strain landscape. The asymmetric distribution of strain is the result of the high tensile strain generated in the elevated regions, such as nanobubbles, compared to the weak compressive strain developed in wrinkles and depressions. This effect has been reproduced successfully in several samples as shown in the SI. Top encapsulation with hBN is not beneficial in these kinds of experiments because it requires applying pressure on the WS$_2$ monolayer that would release most of the strain.

**Photoluminescence and hyperspectral measurement.** Photoluminescence and spectroscopy measurements are carried out in a home-built confocal microscope setup coupled to a closed-cycle cryostat. Except for experiments for resonant excitation that are performed with a supercontinuum pulsed laser, photoluminescence experiments are performed by exciting the samples with a continuous-wave (CW) green laser (532 nm). PL maps are taken with a galvanometer mirror scanner in a 4f configuration. The diffraction-limited spatial resolution is approximately 350 nm. An objective with NA=0.95 and free-space-coupled avalanche photodiodes (APDs) are used for high efficiency collection. In detection, the excitation laser is filtered out with a dichroic mirror and a 550-nm long-pass filter, or with tunable spectral filters to only detect emission from dark excitons. Spectra and hyperspectral maps are obtained by directing the signal to a spectrometer with gratings of either 150 or 600 G/mm. Hyperspectral images are taken by collecting spectra at different locations with steps of 100 nm or 300nm.

**Density functional theory and strain model.** The structural geometry relaxations and the electronic structure calculations under different values of strain are performed using density functional theory. We relax the WS$_2$ structure using ultrasoft pseudopotential with local density approximation (LDA) exchange-correlation and the plane waves implemented in the QUANTUM ESPRESSO package.[36] The momentum space is sampled with 24x24x1 Monkhorst-Pack mesh and the kinetic energy cutoff is set to 70 Ry. The total force on each atom after relaxation is less than 0.0001 Ry/Bohr. We used a vacuum space of 20 Å perpendicular to the WS$_2$ monolayer. The spin-orbit calculation is avoided as it is negligibly affected by strain. We find an optimized lattice constant for the unstrained case of 3.121 Å. The phonon dispersion of WS$_2$ is calculated using the density functional perturbation theory (DFPT) method within the Phonon package of QUANTUM ESPRESSO. We sampled momentum space with an 8x8x1 Monkhorst-Pack mesh grid.

FIGURES

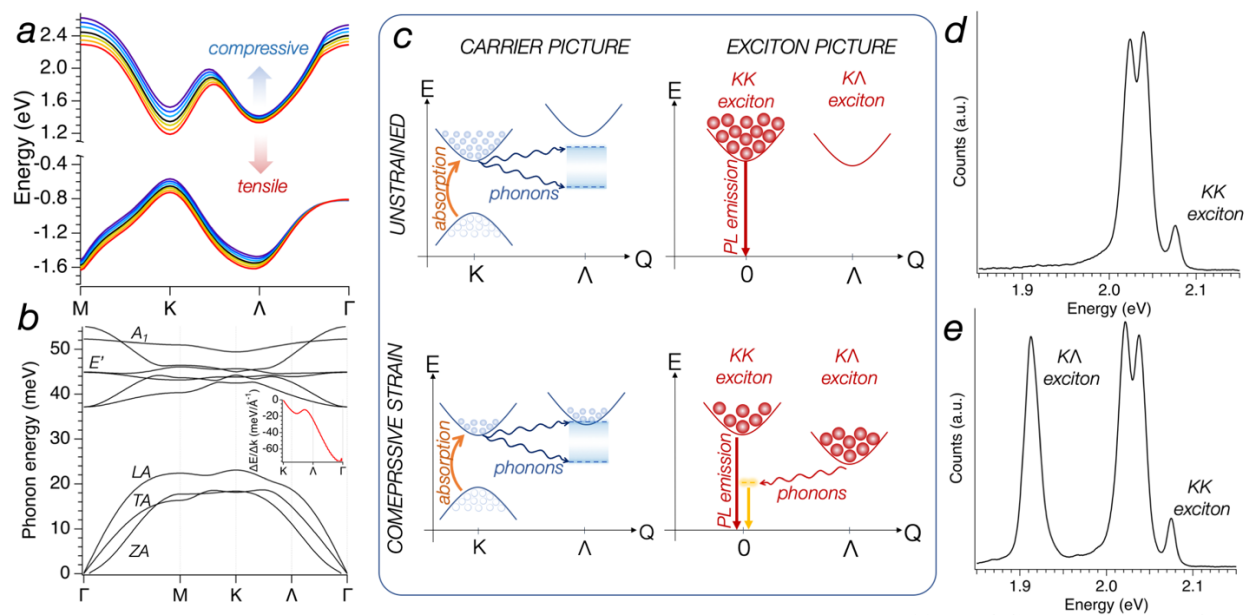

**Figure 1.** Brightening of momentum-forbidden excitons with compressive strain. (a) Band diagram of a monolayer $WS_2$ calculated with DFT methods at different level of strain ranging from -0.6% to +0.6% with steps of 0.2%. The black curve represents the zero-strain dispersion. The valence band (VB) maxima are located at the $\Gamma$ and $K$ high symmetry points, and conduction band (CB) minima at the $\Lambda$ and $K$. The CB maxima at $K$ and $\Lambda$ shift under strain or temperature changes. The energy separation between them ($\Delta E^{K\Lambda}$) reduces when compressive strain is applied. (b) Phononic dispersion of a $WS_2$ monolayer. The inset shows the derivative of the LA phonons around the $\Lambda$ point. (c) Strain-induced formation of $K\Lambda$ excitons. Top (bottom) panel shows a cartoon of the band diagram in the carrier and exciton picture for the unstrained (compressive strain) case. In the unstrained case, electron excitation occurs in the $K$ valley upon optical absorption. Due to the large energy difference between the CB minima, no phonon mode is available to scatter electrons from $K$ to $\Lambda$. In this case, only $KK$ excitons can decay radiatively. When compressive strain is applied (bottom panel) the relative energy difference between the CB minima allows energy and momentum conservation for the electron-phonon scattering process between $K$ and $\Lambda$. Upon excitation to the $K$ valley, electrons can populate the $\Lambda$ valley, forming momentum-forbidden $K\Lambda$ excitons that can then recombine radiatively by phononic emission. (d) and (e) show the emission spectrum of $WS_2$ at T=77 K in the unstrained and compressive strain cases, respectively. When $WS_2$ is compressed a new low-energy peak appears in the spectrum due to the formation of momentum-forbidden excitons.

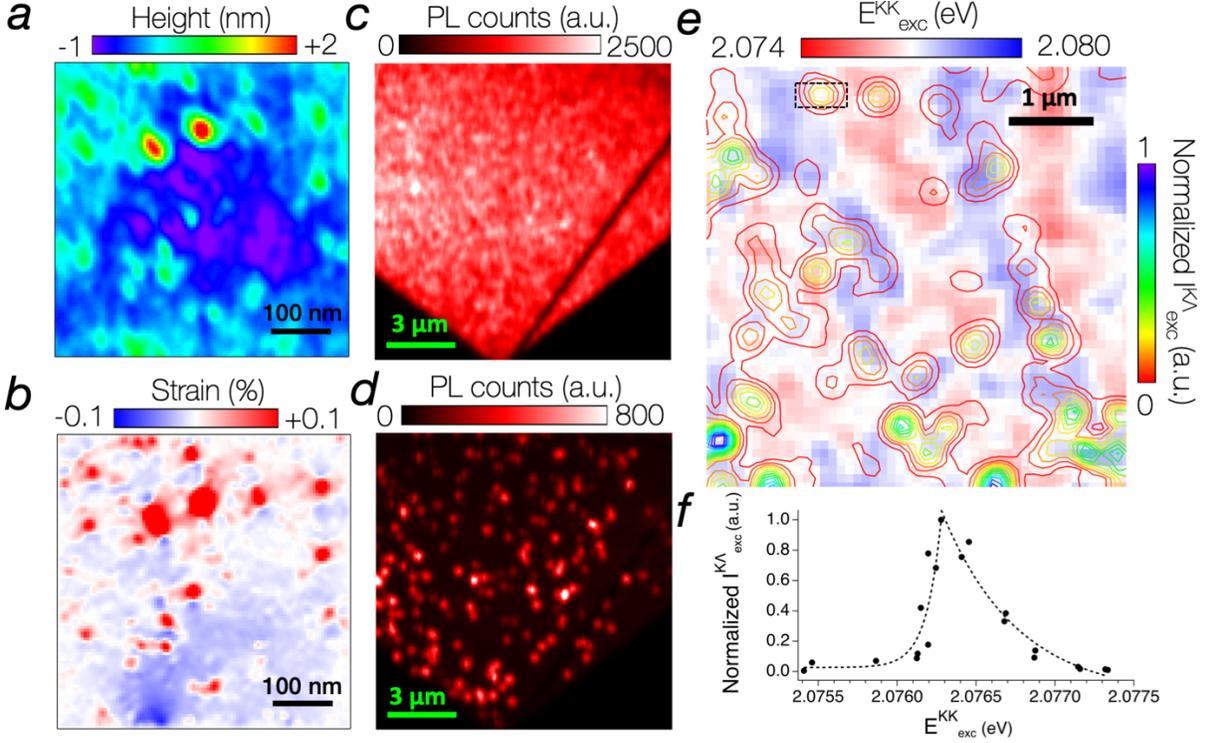

**Figure 2.** Strain-induced formation of dark excitons in $WS_2$. (a) Atomic force microscope topographical scan of 500 nm x 500 nm area of a $WS_2$ monolayer indicates the presence of several nanobubbles and wrinkles in the sample. (b) Strain map obtained from the AFM scan of (a). The strain magnitude ranges from -0.1% to 0.3%. Colormap has been chosen to have a white color associated with regions with zero strain. (c) PL map of a $WS_2$ monolayer at T=77 K. Despite the uneven surface, the PL emission is homogeneous across the sample. (d) Same PL map of (c) but with the signal filtered to collect only photons with E<1.94 eV to include only dark excitons. The emission patterns of the *KΛ* exciton are localized around regions with compressive strain. (e) The colormap shows the variation of the *KK* exciton energy in a small region of the sample. Blueshift corresponds to compressive strain as described in Fig.S1. The counterplot of the *KΛ* exciton emission is overlaid to the *KK* energy map showing that *KΛ* excitons are mostly created in regions with moderated compressive strain. (f) Intensity of the *KΛ* exciton as a function of $E^{KK}_{exc}$ extracted from the region highlighted with the dashed box in (e). The population of KΛ excitons rapidly increases with compressive strain due to the activation of the *K→Λ* phonon scattering process. When compressive strain is increased further, the band alignment reduces the efficiency of the phonon scattering and $I^{KΛ}_{exc}$ slowly decreases. The curves are the result of the fitting with the exponential function $Ae^{-\frac{x-x_0}{B}}$. The fit returns B=-0.9·10⁻⁴ and B=+4.9·10⁻⁴ for the rise and decay of $I^{KΛ}_{exc}$.

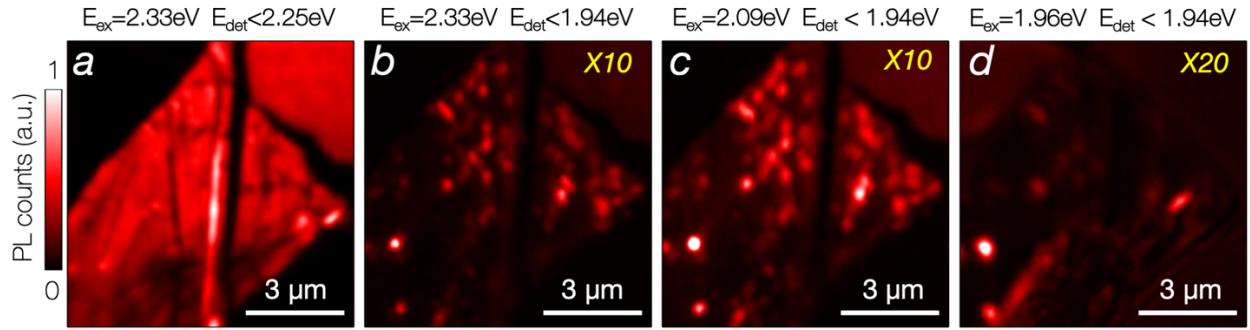

**Figure 3.** Photoluminescence experiments at different excitation energy. (a) PL map of a strained $WS_2$ monolayer at T = 5 K excited with a CW laser at 532 nm (2.331 eV). The laser reflection is filtered out with a long pass filter at 550 nm (2.254 eV). (b) PL map of the same sample of (a) excited with a tunable pulsed laser of 2 nm-bandwidth centered at 532 nm (2.331 eV). The PL emission is filtered with a long pass filter at 640 nm (1.938 eV) in order to collect only emission from dark excitons. For the sake of visibility, the colormap is rescaled by a factor 10. (c) PL map of the same sample excited with a tunable pulsed laser at 592 nm (2.094 eV) in resonance with the *KK* exciton. The PL emission is filtered at 640 nm (1.938 eV) in order to collect only emission from dark excitons. For the sake of visibility, the colormap is rescaled by a factor 10. (d) PL map of the same sample excited with a tunable pulsed laser at 632 nm (1.962 eV) below the resonance of the *KK* exciton. The PL emission is filtered at 640nm (1.938 eV) in order to collect only emission from dark excitons. For the sake of visibility, the colormap is rescaled by a factor 20.

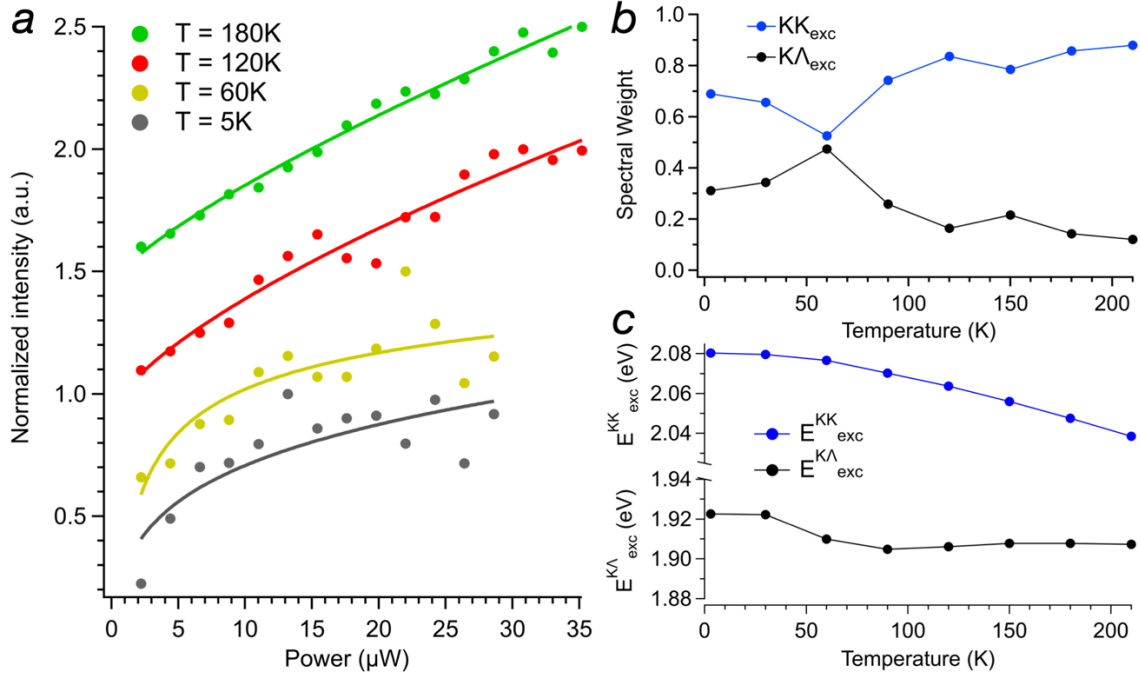

**Figure 4.** Dark exciton emission at different temperature and excitation power. (a) Normalized emission intensity of the *KΛ* exciton as a function of the laser pump power at different temperature displayed with a vertical offset of 0.5. Data are normalized to the maximum intensity for each temperature. Curves are the result of the fitting with a power function: $I(P)=AP^w$, where A is a fitting constant. At low temperature, the *KΛ* exciton quickly saturates with a power exponent as low as *w*=0.18. The saturation of the intensity is the result of the low density of state of LA phonons responsible for the formation of the momentum-forbidden excitons. When temperature is increased, more phonons are available and the intensity shows an almost linear dependency on the pump power. (b) Spectral weight of the *KK* and *KΛ* excitons as a function of temperature showing a peak of efficiency in the formation of momentum-forbidden excitons around T=60 K. (c) Emission energy of both *KK* and *KΛ* excitons as a function of temperature. While the *KK* exciton shows a redshift of about 40 meV, the energy of the *KΛ* exciton only changes about 15 meV.

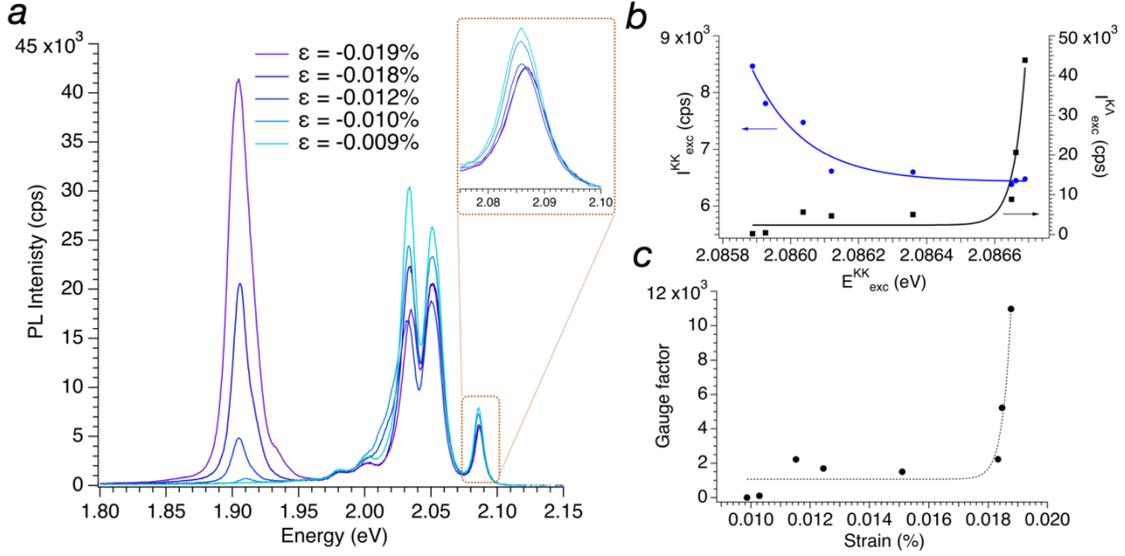

**Figure 5. Strain sensing with dark excitons.** (a) Emission spectra at different compressive strain levels at T=60 K. Spectra are extracted from the hyperspectral map in a location characterized by compressive strain, as described in the SI. The increase of compression in the WS$_2$ monolayer is characterized by a rise of the *KΛ* exciton peak as well as a blueshift of the *KK* excitons. (b) Emission intensity of the *KK* exciton ($I^{KK}_{exc}$ in blue dots) and *KΛ* exciton ($I^{KΛ}_{exc}$ in black dots) as a function of the energy of the *KK* exciton. The curves show the result of the fitting with an exponential function. The exponential decrease (increase) of $I^{KK}_{exc}$ ($I^{KΛ}_{exc}$) is the result of the activation of the *K→Λ* scattering channel and the formation of momentum-forbidden excitons. (c) The gauge factor shows a nonlinear dependency on the strain with a maximum approaching 12000.

## ASSOCIATED CONTENT

**Supporting information.** The supporting information is available free of charge: DFT simulations of Electron, phonon, and exciton energies; Further sample preparation details; AFM and PL maps of strained $WS_2$ monolayers; Resonant excitation experiment; Hyperspectral analysis details; Calculation of strain from AFM topography; and T=5 K spectral analysis.

## AUTHOR INFORMATION

**Corresponding Author**

*Address correspondence to ggrosso@gc.cuny.edu

**Author Contributions**

S.B.C. and G.G. conceived the idea/concept and defined the experimental and theoretical work. S.B.C. and J.M.W. prepared samples and performed experimental measurements. S.B.C. performed DFT simulations and strain calculations. G.G. and S.B.C. analyzed the data. S.B.C., G.G., J.M.W. and E.M. participated in discussions and the writing of the manuscript. T.T. and K.W grew hBN samples. G.G. conceived and directed the project.

**Notes**

The authors declare no competing financial interest.

## ACKNOWLEDGMENTS

G.G. acknowledges support from the National Science Foundation (NSF) (grant no. DMR-2044281), support from the physics department of the Graduate Center of CUNY and the Advanced Science Research Center through the start-up grant, and support from the Research Foundation through PSC-CUNY award 64510-00 52. K.W. and T.T. acknowledge support from the Elemental Strategy Initiative conducted by the MEXT, Japan (Grant Number JPMXP0112101001) and JSPS KAKENHI (Grant Numbers 19H05790, 20H00354 and 21H05233).